\documentclass[12pt,aps,prl,showkeys,showpacs,,longbibliography,preprint ]{revtex4-1}   

\usepackage{amsmath}    
\usepackage{amsfonts}  
\usepackage{amssymb}
\usepackage{mathrsfs}
\usepackage{graphicx}   
\usepackage{epstopdf}

\newcommand{\rd}{{\mathrm d}}

\begin{document}
%
\title{On the metal-insulator-transition in vanadium dioxide}
%

\author{Shigeji Fujita,$^1$ Azita Jovaini,$^1$ Salvador Godoy,$^2$ and Akira Suzuki$^{3*}$}
\affiliation{$^1$Department of Physics, University at Buffalo, SUNY, Buffalo, New York 14260, USA,}

\affiliation{$^2$Departamento de F\'{\i}sica, Facultad de Ciencias, Universidad Nacional Aut\'{o}noma de M\'{e}xico, M\'{e}xico D.F.  04510, M\'{e}xico}

\affiliation{$^3$Department~of~Physics, Faculty~of~Science, Tokyo~University~of~Science, Shinjyuku-ku, Tokyo~162-8601, Japan}
\email{asuzuki@rs.kagu.tus.ac.jp}
\date{\today}

\begin{abstract}
Vanadium dioxide (VO$_2$) undergoes a metal-insulator transition (MIT) at 340 K with the structural change between tetragonal and monoclinic crystals as the temperature is lowered.  The conductivity $\sigma$ drops at MIT by four orders of magnitude.  The low-temperature monoclinic phase is known to have a lower ground-state energy.  The existence of a $k$-vector ${\boldsymbol k}$ is prerequisite for the conduction since the ${\boldsymbol k}$ appears in the semiclassical equation of motion for the conduction electron (wave packet).  Each wave packet is, by assumption, composed of the plane waves proceeding in the ${\boldsymbol k}$ direction perpendicular to the plane.  The tetragonal (VO$_2$)$_3$ unit cells are periodic along the crystal's $x$-, $y$-, and $z$-axes, and hence there are three-dimensional $k$-vectors.  The periodicity using the non-orthogonal bases does not legitimize the electron dynamics in solids.
There are one-dimensional ${\boldsymbol k}$ along the $c$-axis for a monoclinic crystal.  We believe this decrease in the dimensionality of the $k$-vectors is the cause of the conductivity drop. Triclinic and trigonal (rhombohedral) crystals have no $k$-vectors, and hence they must be insulators.  The majority carriers in graphite are ``electrons", which is shown by using an orthogonal unit cell for the hexagonal lattice. 
\end{abstract}
%
\maketitle
\begin{figure}[htbp]
\centering
\includegraphics[scale=0.45]{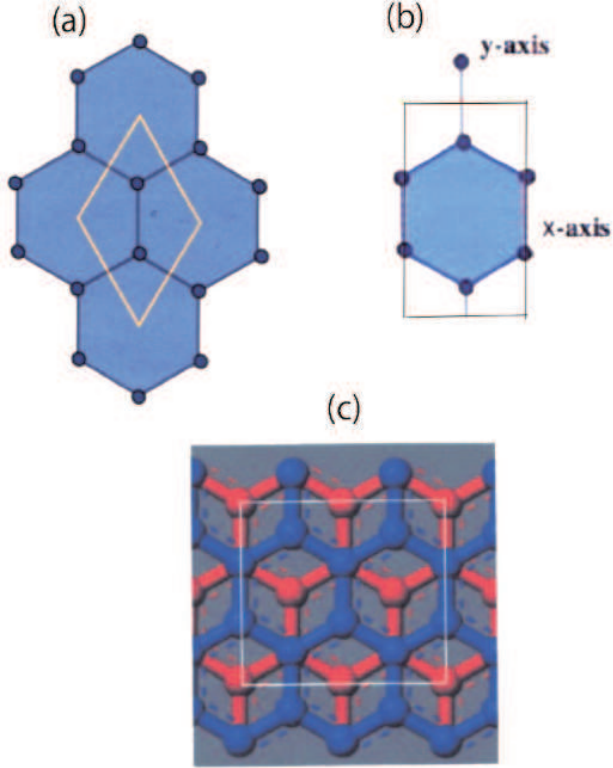}
\caption{(a) WS unit cell, rhombus ({yellow solid lines}) for graphene.  (b) The orthogonal unit cell, rectangle (black solid lines).  (c) An orthogonal unit cell ({white solid lines}) viewed from the top.  The carbons (circles) in the A (B) planes are shown in blue (orange).}\label{fig1}
\end{figure}

%
%
{\em{Introduction}} -- In 1959 Morin reported his discovery of a metal-insulator  transition (MIT) in vanadium dioxide (VO$_2$)~\cite{Morin}.  Compound VO$_2$ forms a monoclinic (MCL) crystal on the low temperature side and  a tetragonal (TET) crystal on the high temperature side.  When heated, VO$_2$ undergoes an insulator-to-metal transition around 340 K, with the resistance drop by four orders of magnitude.  The phase change carries a hysteresis just as a ferro-to-paramagnetic phase change.  The origin of the phase transition has been attributed by some authors to Peierls instability driven by strong electron--phonon interaction~\cite{Goodenough}, and by other authors to Coulomb repulsion and electron localization due to the electron-electron interaction on a Mott--Hubbard picture~\cite{Mott,Wentzcovitch,Rice}.\\
\indent
A simpler view on the MIT is presented in this Letter.  The MLC phase is known to have a lower ground-state energy than the TET phase. The existence of $k$-vectors is prerequisite for the electrical conduction since the ${\boldsymbol k}$ appear in the semiclassical equation of motion for the conduction electron (wave packet).  The TET (VO$_2$)$_3$ unit cells are periodic along the crystal's $x$-, $y$-, and $z$-axes, and hence there are three--dimensional (3D) $k$-vectors.  There are 1D $k$-vectors along the $c$-axis for a MCL crystal.  We show that the MIT occurs since the dimensionality of the $k$-vectors is reduced from three (3) to one (1) in going from the TET to the MCL crystals.  We also show that the majority carriers in graphite are ``electrons" based on the orthogonal unit cells for the hexagonal crystals.  In this letter conduction electrons are denoted by quotation marked ``electrons" (``holes") whereas generic electrons are denoted without quotation marks.  \\
%
%
\indent
{\em{Theory}} -- Following Ashcroft and Mermin~\cite{AM}, we adopt the semiclassical model of electron dynamics in solids.  It is necessary to introduce $k$-vectors:
\begin{equation}
{\boldsymbol k}=k_x\widehat{\mathbf e}_x+k_y\widehat{\mathbf e}_y+k_z\widehat{\mathbf e}_z\,, \label{1}
\end{equation}
where $\widehat{\mathbf e}_x$, $\widehat{\mathbf e}_y$, $\widehat{\mathbf e}_z$ are the orthonormal unit vectors, since the ${k}$-vectors are involved in the semiclassical equation of motion:
\begin{equation}
\hbar\dot{\boldsymbol k} \equiv \hbar\frac{\rd{\boldsymbol k}}{\rd t}=q({\boldsymbol E} + {\boldsymbol v}\times{\boldsymbol B}),\label{2}
\end{equation}
where $q$ is the charge of a conduction electron, and ${\boldsymbol E}$ and ${\boldsymbol B}$ are the electric and magnetic fields,  respectively.  The vector
\begin{equation}
{\boldsymbol v}\equiv\frac{1}{\hbar}\frac{\partial\varepsilon}{\partial{\boldsymbol k}}\label{3}
\end{equation}
is the electron velocity, where $\varepsilon = \varepsilon({\boldsymbol k})$ is the energy. \\
\indent
If we introduce the mass tensor ${\mathcal M}$ defined by
\begin{equation}
[{\mathcal M}^{-1}]_{ij}\equiv\frac{1}{\hbar^2}\frac{\partial^2\varepsilon}{\partial k_i\partial k_j}\,,\label{4}
\end{equation}
then the equations of motion can be written as~\cite{Fujita}
\begin{equation}
\sum_jm_{ij}\frac{\rd v_j}{\rd t} = q({\boldsymbol E} + {\boldsymbol v} \times {\boldsymbol B})_i\,.\label{5}
\end{equation}
The mass tensor ${\mathcal M}$ is symmetric:
\begin{equation}
m_{ij}=m_{ji}\,, \label{6}
\end{equation}
and can be characterized by the effective masses $\{m_{i}^*\}$.  If we choose a Cartesian coordinate system along the principal axes of the mass tensor, we can write  Eq.~(\ref{5}) as~\cite{Fujita}
\begin{equation}
m_j^*\frac{\rd v_j}{\rd t}=q({\boldsymbol E} + {\boldsymbol v}\times{\boldsymbol B})_j \,.\label{7}
\end{equation}
In this form the Newtonian character of the equations of motion is transparent.  If an electron is in a continuous energy range (energy band), then it will be accelerated by the electric force $q{\boldsymbol E}$ following Eq.~(\ref{7}), and the material is a conductor.  If the electron's energy is discrete and is in a forbidden band (energy gap), it does not move under a small electric force, and the material is insulator.  If the acceleration occurs only for a mean free time (the inverse of a scattering frequency) $\tau$, the conductivity $\sigma$ for a simple metal is given by  Drude's formula~\cite{AM}:
\begin{equation}
\sigma=q^2n\tau/m^*\,,\label{8}
\end{equation}
where $n$ is the electron density and $m^*$ the effective mass.\\
\indent 
For some crystals such as simple cubic (SC), face-centered cubic (FCC), body-centered-cubic (BCC), tetragonal (TET) and orthorhombic (ORC) crystals, the choice of the orthogonal $(x,y,z)$-axes and the unit cells are obvious.  The 2D crystals can also be treated similarly, only the $z$-component being dropped.\\
\indent
We assume that the wave packet is composed of superposable plane-waves characterized by the $k$-vectors.  The superposability is the basic property of the Schr\"odingier wave function in free space.  A MCL crystal can be generated from an ORC crystal by distorting the rectangular faces perpendicular to the $c$-axis into parallelograms.  Material plane-waves proceeding along the $c$-axis exist since the $(x, y)$ planes containing materials (atoms) are periodic in the $z$-direction in equilibrium and can execute small oscillations.  It has then one-dimensional (1D) $k$-vectors along the $c$-axis.  In the $x$-$y$ plane there is an oblique net whose corners are occupied by V's for MCL VO$_2$.  The position vector ${\boldsymbol R}$ of every V can be represented by integers $(m,n)$, if we choose
\begin{equation}
{\boldsymbol R}_{mn}=m{\boldsymbol a}_1 + n{\boldsymbol a}_2\,,\label{9}
\end{equation}
where ${\boldsymbol a}_1$ and ${\boldsymbol a}_2$ are non-orthogonal base vectors.  In the field theoretical formulation the field point ${\boldsymbol r}$ is given by
\begin{equation}
{\boldsymbol r}={\boldsymbol r}^\prime + {\boldsymbol R}_{mn}\,,\label{10}
\end{equation}
where ${\boldsymbol r}^\prime$ is the point defined within the standard unit cell.  
Eq.~(\ref{10}) describes  the 2D lattice periodicity but does {\emph{not}} establish $k$-space as explained below.\\
\indent
To see this clearly, we first consider an electron in a simple square (sq) lattice.  The Schr\"odingier wave equation is
\begin {equation}
i\hbar\frac{\partial}{\partial t}\psi({\boldsymbol r})=-\frac{\hbar^2}{2m^*}\nabla^2\psi({\boldsymbol r})+V({\boldsymbol r})\psi({\boldsymbol r})\,,\label{11}
\end{equation}
where the potential energy $V$ is periodic:
\begin{align}
&V({\boldsymbol r}+{\boldsymbol R}_{mn}^{(0)})=V({\boldsymbol r})\,,\label{12}\\
&{\boldsymbol R}_{mn}^{(0)}\equiv m{\boldsymbol a}_x + n{\boldsymbol a}_y = ma{\widehat{\mathbf e}}_x + na{\widehat{\mathbf e}}_y\,,\label{13}\\
&(a = \hbox{lattice constant})\,. \nonumber
\end{align}
If we choose a set of Cartesian coordinates $(x, y)$ along the sq lattice, then the Laplacian term in Eq.~(\ref{11}) is given by
\begin{equation}
\nabla^2\psi (x,y)=\left(\frac{\partial^2}{\partial x^2}+\frac{\partial^2}{\partial y^2}\right)\psi(x,y)\,.\label{14}
\end{equation}
If we choose a periodic square boundary with the side length $Na$, $N=$integer, then there are 2D Fourier transforms and (2D) $k$-vectors.
\indent
We now go back to the original rhombic system.  If we choose the $x$-axis along either ${\boldsymbol a}_1$ or ${\boldsymbol a}_2$, then the potential energy field $V({\boldsymbol r})=V({\boldsymbol r}+{\boldsymbol R}_{mn})$ is periodic in the $x$-direction but it is aperiodic in the $y$-direction.  We can obtain no 2D periodic boundary condition suitable for Fourier transformation.  Then, there is no 2D $k$-space.  If we omit the kinetic energy term, then we can still use Eq.~(\ref{9}) and obtain the ground state energy (except the zero point energy).  The reduction in the dimensionality of the $k$-vectors from 3D to 1D is the cause of the conductivity drop in the MIT.  The MIT proceeds by domains since the insulator (MCL) phase has the lower degrees of symmetry.  Strictly speaking, the existence of 1D $k$-vectors allows the MCL material to have a small conductivity.  This happens for VO$_2$, see below.
\indent
Wu {\em et al}.~\cite{Wu} measured the resistance $R$ of individual nanowires W$_x$V$_{1-x}$O$_x$ with tungsten (W) ranging up to 1.14 $\%$.  The nanowires are grown with the wire axis matching with the $c$-axis of the high-temperature rutile structure.  The transition temperature $T_0$ decreases from 340 K, passing the room temperature, to 296 K as the concentration $x$ changes from 0 to 1.14 $\%$.  The temperature dependence of $R$ for the low temperature phase is semiconductor-like.  That is, the resistance $R$ decreases with increasing temperature.  The behavior can be fit with the Arrhenius law:
\begin{equation}
\sigma \propto R^{-1} \propto \exp(-\varepsilon_{\mathrm a}/k_{\mathrm B}T)\,, \label{15}
\end{equation}
where $\varepsilon_{\mathrm a}$ is the activation energy and $k_{\mathrm B}$ the Boltzmann constant.  
\indent
There are seven (7) crystal systems, as seen in AM's book~\cite{AM}.  They are: cubic (CUB), TET, ORC, MCL, rhombohedral (RHL), hexagonal (HEX), and triclinic systems.  There are no $k$-vectors for triclinic and RHL systems.  Materials forming these crystals must be true insulators.   Arsenic (As) and Bismuth (Bi) form RHL crystals, and they are insulators.  Diamond (C), silicon (Si) and germanium (Ge) form diamond (DIA) crystals.  A DIA lattice can be decomposed into two FCC sublattices, and can therefore be treated similarly to a CUB crystal.  A sizable number of elements including graphite form HEX crystals.  HEX crystals can be treated similar to ORC crystals by choosing orthogonal unit cells.  See below for the case of graphite.\\
%
%
\indent
{\em{Graphene}} -- We consider a graphene which forms a 2D honeycomb lattice.  The Wigner-Seitz (WS) unit cell, a rhombus ({yellow lines}) shown in Fig.~\ref{fig1}\,(a), contains 2 C's.  We showed in our earlier work~\cite{Suzuki} that graphene has ``electrons" and ``holes" based on the rectangular unit cell (black solid lines) shown in Fig.~\ref{fig1}\,(b).  We briefly review our calculations.  More details can be found in Refs.~\cite{Fujita3,Suzuki}.
%
%
\indent We assume that the ``electron" (``hole") wave packet has the charge $-e$ $(+e)$ and a size of the rectangular unit cell, generated above (below) the Fermi energy $\varepsilon_{\mathrm F}$.  We could show~\cite{Suzuki} that (a) the ``electron" and the ``hole" have different charge distributions and different effective masses, (b) that the ``electrons" and ``holes" move in different easy channels, (c) that the ``electrons" and ``holes" are thermally excited with different activation energies, and (d) that the ``electron" activation energy $\varepsilon_1$ is smaller than the ``hole" activation energy $\varepsilon_2$:
\begin{equation}
\varepsilon_1<\varepsilon_2\,.\label{16}
\end{equation}
Thus, ``electrons'' are the majority carriers in graphene.  The thermally activated electron densities are then given by
\begin{equation}
n_j(T)=n_je^{-\varepsilon_j/k_{\mathrm B}T}\,,\quad n_j=\hbox{constant},\label{17}
\end{equation}
where $j=1$ and $2$ represent the ``electron'' and ``hole'', respectively.  Magnetotransport experiments by Zhang {\em et al}.~\cite{Zhang} indicate that the ``electrons" are majority carriers in graphene.  Thus, our theory is in agreement with experiments.  \\ 
%
%
\indent
{\em{Graphite}} -- Graphite is composed of graphene layers stacked in the manner ABAB$\cdots$ along the $c$-axis.  We may choose an orthogonal unit cell shown in Fig.~\ref{fig1}\,(c).  The unit cell contains 16 C's.  The two rectangles (white solid lines) are stacked vertically with the interlayer  separation, $c_0=3.35$ \AA\,\,\,much greater than the nearest neighbor distance between two C's, $a_0=1.42 $ \AA:
\begin{equation}
c_0\gg a_0\,.\label{18}
\end{equation}
The unit cell has three side-lengths:
\begin{equation}
b_1=3a_0,\quad b_2=2\sqrt3a_0,  \quad  b_3=2c_0\,. \label{19}
\end{equation}
Clearly, the system is periodic along the orthogonal directions with the three periods $(b_1, b_2, b_3)$ given in Eq.~(\ref{19}).  We assume that both ``electron'' and ``hole'' have the same unit cell size.  In summary the system is orthorhombic with the sides $(b_1, b_2, b_3)$, $b_1\ne b_2$, $b_1\ne b_3$, $b_2\ne b_3$.

The negatively charged ``electron'' (with the charge $-e$) in graphite are welcomed by the positively charged C$^+$ when moving vertically up or downwards.  That is, the easy directions for the ``electrons'' are vertical.  The easy directions for the ``holes'' are horizontal.  There are no hindering hills for ``holes'' moving horizontally.  Hence, the ``electron'' in graphite has the lower activation energy $\varepsilon$ than the ``hole'':  
$\varepsilon_1 < \varepsilon_2$.
Then, ``electrons''  are the majority carriers in graphite.  The thermoelectric power (Seebeck coefficient) measurements by Kang {\em et al}.~\cite{Kang} show that the majority carriers in graphite are ``electrons", which is in agreement with our theory.

It is often said~\cite{Saito} that since the separation distance $c_0$ is much greater than the nearest neighbor distance $a_0$, the conduction in graphite is two-dimensional, and can be discussed in terms of the motion in the graphene as a first approximation.  We take a different point of view.  The conduction electrons move as wave packets having the 3D orthogonal unit cell sizes.  The conduction is two-dimensional because of the inequality (\ref{18}).  But the transport behaviors in graphite and graphene are very different because of the different unit cells.

The construction of the orthogonal unit cell developed here can be followed in other material forming HEX crystals.  Zinc (Zn) and Beryllium (Be) form HEX crystals.  The closed orbits on the coronet-like Fermi surface~\cite{AM} generate cyclotron resonance, which may be discussed using the orthogonal unit cells.

In summary we established that
(a) The MIT in VO$_2$ directly arises from the lattice structure change between the TET and the MCL crystals.  The TET (MCL) crystal has 3D (1D) ${\boldsymbol k}$-vectors.  The reduction in the dimensionality of the $k$-vectors causes a conductivity drop,
(b) Triclinic and trigonal crystals have no $k$-vectors, and hence they are insulators,
(c) Hex graphite has ``electrons" as the majority carriers. \\
\indent
{\em{Acknowledgments}}:  The authors thank Professor S.~Banerjee and Professor P.~Zhang for enlightening discussions.
%

%
\end{document}